\documentclass[%
 reprint,
superscriptaddress,
 amsmath,amssymb,
 aps,
longbibliography
]{revtex4-2}

\usepackage{graphicx}% Include figure files
\usepackage{dcolumn}% Align table columns on decimal point
\usepackage{bm}% bold math
\usepackage{multirow}
\usepackage{soul}
\usepackage[dvipsnames]{xcolor}
\usepackage{xfrac}
\usepackage[colorlinks,plainpages=false,linkcolor=blue,urlcolor=blue,citecolor=blue,pdfpagemode=UseNone]{hyperref}
\usepackage{tikz}
\usetikzlibrary{quantikz}
\usepackage{adjustbox}

\newcommand{\figrefx}[2]{\hyperref[#1]{\ref*{#1}(#2)}}
\newcommand{\figrefxx}[5]{\hyperref[#1]{\ref*{#1}(#2),(#3),(#4),(#5)}}
\newcommand{\figrefxxx}[6]{\hyperref[#1]{\ref*{#1}(#2),(#3),(#4),(#5),(#6)}}
\newcommand{\figrefxy}[3]{\hyperref[#1]{\ref*{#1}(#2)--\ref*{#1}(#3)}}

\newcommand{\etal}{{\sl et al.}}
\newcommand{\ie}{i.e.}
\newcommand{\alo}{$\alpha$-Al$_2$O$_3$ (0001)}
\newcommand{\Heff}{$H_{\mathrm{eff}}$}
\newcommand{\mycomment}[1]{}                                      %%% To change the text into comment

\usepackage{lineno}
\newcommand{\beginsupplement}{
        \setcounter{table}{0}
        \renewcommand{\thetable}{S\arabic{table}}
        \setcounter{figure}{0}
        \renewcommand{\thefigure}{S\arabic{figure}}
        \setcounter{equation}{0}
        \renewcommand{\theequation}{S\arabic{equation}}
        \setcounter{page}{1} % Reset page counter to 1
        \renewcommand{\thepage}{S\arabic{page}} % Change page numbering style for supplementary material
        \onecolumngrid
}

\newcounter{Suppfigcounter}
\newenvironment{Suppfig}{\refstepcounter{Suppfigcounter}}{}
\newcommand{\supref}[1]{\hyperref[#1]{S\ref*{#1}}}

\begin{document}

\preprint{APS/123-QED}
\title{Quantum simulations of defects near the (0001) surface of $\alpha$-Al$_2$O$_3$}
%Limit of 8 lines

\author{Vijaya Begum-Hudde}
%\email{vijayab@illinois.edu}
\affiliation{Department of Materials Science and Engineering, University of Illinois Urbana-Champaign, Urbana, IL 61801, USA}

\author{Yi-Ting Lee}
\affiliation{Department of Materials Science and Engineering, University of Illinois Urbana-Champaign, Urbana, IL 61801, USA}

\author{Barbara A.\ Jones}
\thanks{manuscript submitted posthumously}
\affiliation{IBM Research Almaden Lab, 650 Harry Rd, San Jose, CA 95120}

\author{Andr{\'e} Schleife}
\email{schleife@illinois.edu}
\affiliation{Department of Materials Science and Engineering, University of Illinois Urbana-Champaign, Urbana, IL 61801, USA}
\affiliation{Materials Research Laboratory, University of Illinois Urbana-Champaign, Urbana, IL 61801, USA}
\affiliation{National Center for Supercomputing Applications, University of Illinois Urbana-Champaign, Urbana, IL 61801, USA}

\date{\today}

\begin{abstract}
Defects in materials are ubiquitous and one of their adverse effects in $\alpha$-Al$_2$O$_3$ is the initiation of corrosion.
While this process starts near the surface, the defects involved and their electronic structure need to be elucidated with high accuracy.
Since point defects are confined to a small spatial region, defect embedding theory allows the definition of an active space, comprising of the defect electronic states, that is coupled to the environment of the host material.
The active space Hamiltonian is of small rank, enabling access to its electronic properties using a high-level or even exact quantum theory.
In this paper we use these techniques and first-principles simulations to compute the structural and electronic properties of near-surface vacancies for the (0001) surface of $\alpha$-Al$_2$O$_3$, and investigate the influence of defects and hydration on the initiation and propagation of corrosion.
We report the defect electronic structure for strongly localized ground and excited states of the surface O vacancy and compare results obtained using full configuration interaction and a variational quantum eigensolver on a quantum computer.
Error mitigation techniques are explored and shown to reduce the error due to the hardware noise to the point where the quantum result agrees with the exact solution within chemical accuracy.
\end{abstract}

\maketitle

\section{Introduction}

Having entered the noisy intermediate-scale quantum (NISQ) era \cite{Preskill-2018}, quantum computers are becoming accessible for exploring their use to solve a wide range of scientific problems.
As one example, IBM currently offers hardware architectures with 127 qubits for quantum simulations, and recently, a 1,121-qubit superconducting qubit quantum processor has been announced~\cite{Castelvecchi-2023}.
%(https://www.nature.com/articles/d41586-023-03854-1)
Applications of such systems include the determination of the ground-state energy e.g.\ of He-H$^+$ \cite{Peruzzo-2014}, BeH$_2$\cite{Kandala-2017}, and H$_{12}$ \cite{Arute-2020} on two, six, and twelve qubits, respectively.
To achieve quantum advantage over classical computers, algorithms to solve more complex problems in condensed-matter physics, including error mitigation, are needed.

In current quantum hardware various types of noise are prevalent arising from hardware imperfections, thermal noise, and dephasing in qubits~\cite{Johnstun-2021}.
These have a negative impact on qubit functional parameters such as coherence time and fidelity.
Hence, in order to fully harness the capabilities of NISQ quantum computers, it is critical to address noise arising from the hardware, e.g.\ via fabrication of quantum devices with high fidelity \cite{Leon-2021,Castelvecchi-2023} or through error correction \cite{Steane-1996,Ryan-Anderson-2021}.
In the meantime, various error mitigation techniques exist, including zero noise extrapolation (ZNE)~\cite{Temme-2017,Li-PRX-2017,Kandala-2019}, probabilistic error correction~\cite{Temme-2017,Endo-2018,Zhang-Nature-2020}, stochastic error mitigation~\cite{Sun-2021}, and new methods are continually being explored.

The straightforward approach of ZNE~\cite{Li-PRX-2017,Dumitrescu-2018,Kandala-2019,Fauseweh-2021} exploits the fact that noise scales with the number of gates: often the two-qubit CNOT gate is considered as a significant source of noise~\cite{Li-PRX-2017}.
For shallow depth circuits, this approach is efficient and was successfully used for error mitigation \cite{Li-PRX-2017,Dumitrescu-2018,Kandala-2019}.
However, for a circuit with a considerable number of CNOT gates, the noise becomes dominant, rendering a barren plateau with vanishing gradient~\cite{McClean-2018} that makes extrapolation to zero noise ineffective. 
Hence, in general the error mitigation technique needs to be chosen specifically for a given Hamiltonian and quantum algorithm.
In this paper, we explore these questions for a technologically important material, Al$_2$O$_3$, and for the specific case of IBM quantum hardware.

Alumina, Al$_2$O$_3$, is employed in various applications such as heterogeneous catalysis and superconducting quantum Josephson junctions, among others~\cite{Yang-2023,Tangcharoen-2023,Makhlin-2001,Clarke-2008,Hutchings-2017}.
It is of interest to the aerospace industry as a thin oxide scale that protects aluminum parts from corrosion by slowing oxygen diffusion from the environment to the underlying aluminum \cite{Zhu-2024}.
The diffusion kinetics are determined by defects in alumina as well as their diffusion barriers and migration pathways, requiring a comprehensive understanding of defects in this material.
In bulk $\alpha$-Al$_2$O$_3$~\cite{Kononov-2023}, a wide array of unconventional, previously unknown native defect configurations were observed, but their role for oxygen diffusion remains unexplored. 

Especially near the surface of $\alpha$-Al$_2$O$_3$, defects can be detrimental for corrosion protection, because they are the starting point of the oxide scale deterioration.
Understanding of surface and interface defects is also required for near-surface qubits \cite{Anderson-2023} where such defects limit the device performance, e.g.\ at the Al/Al$_2$O$_3$ interface \cite{Muller-2019}.
Because it is impossible to eliminate these defects entirely, theoretical modeling of the electronic structure of the defect is required to predict device efficiency \cite{Holder-2013,Gordon-2014,Muller-2019} and devise routes for defect passivation.

Experimentally, non-destructive techniques are available to probe the defect electronic structure and characterize their spectral signatures. 
Spectroscopy involves the excitation of valence and core electrons, and provides information on the excitation energies and local defect environment. 
Since defects modify the electronic structure, changes in the charge state and local environment are captured in the excitation energies, and can be used to identify defects \cite{Katayama-2021}. 

To achieve an accurate description of defect electronic properties, standard density functional theory (DFT) is insufficient and more accurate approaches to describe exchange and correlation are needed.
For example, a previous study shows accurate predictions for defects in bulk Al$_2$O$_3$ based on a hybrid exchange-correlation functional~\cite{Gordon-2014}.
Such simulations have not been performed for the surface and its defects.
Other accurate electronic-structure approaches include many-body perturbation theory and quantum chemical techniques such as configuration interaction and these also provide access to excited-state properties.
However, they are computationally expensive even for bulk systems and approximations are required for defects or surfaces, since these are studied using large supercells.

To compute ground and excited-state properties of  only a small number of defect states, defect embedding theory is a reliable approach.
It can be combined with highly accurate or exact simulations, allowing, for instance, addressing the multireference nature of spin states that cannot be described within conventional mean-field theory~\cite{Cohen-2012}.
Various defect embedding approaches have emerged \cite{Sun-2016}, including techniques based on the density matrix~\cite{Knizia-2012,Kinizia-2013,Wouters-2016,Pham-2020,He-2022}, the electron density~\cite{Huang-2006,Huang-2011,Goodpaster-2014,Genova-2014,Jacob-2014,Wesolowski-2015}, and Green's functions~\cite{Inglesfield-1981,Aryasetiawan-2004,Aryasetiawan-2009,Imada-2010,Dvorak-2019,Zhu-2019}.

Quantum defect embedding theory (QDET) was used to study strongly correlated defect states in materials, and was successfully employed for spin defects in solids such as the negatively charged nitrogen-vacancy (NV$^{-}$) center and neutral silicon-vacancy (SiV$^{0}$) in diamond \cite{Ma-Nature-2020,Ma-2021,Nan-2022}. 
In this approach, starting from the mean-field solution for a large system with a defect, the localized Kohn-Sham (KS) states near the defect are identified and an active space is defined consisting of these states. 
The remainder of the KS states are treated as the environment. 
An effective Hamiltonian for the active space is then constructed to incorporate Coulomb interactions within the localized states and the effect of environment on the active space.
The ground state of a small rank effective Hamiltonian can be solved using the variational quantum eigensolver (VQE)~\cite{Peruzzo-2014}.
This has been shown to be successful for calculating electronic properties for systems up to a dozen atoms~\cite{Shen-2017,Kandala-2017,Ryabinkin-2018,Google-2020,Gao-2021,Rossmannek-2021,Kawashima-2021,Gujarati-2023}.

Interestingly, due to the contraction of the many-body problem to only a few electrons, QDET is compatible with implementation on NISQ computers. 
Previously, this was achieved for solids, using an active space comprising of a few electrons \cite{Gujarati-2023,Huang-2022,Huang-2023}. 
Both ground- and excited-state properties were evaluated on a quantum computer for the spin defects in diamond and SiC using the VQE and sub-space search VQE~\cite{Huang-2022,Huang-2023} along with error mitigation to reduce noise from the hardware.

In this work, we study the role of near-surface defects in corrosion, and successfully solve a minimum QDET model for the near-surface O vacancy in Al$_2$O$_3$ on a quantum computer.
Using first-principles calculations we show that the near-surface O vacancy is more likely to form than an Al vacancy, and, thus, the formation of a pit, comprising of an O and Al vacancy, is most likely initiated at O vacancies.
Hydration of near-surface vacancies is simulated by adsorption of an O adatom and we find that mono O vacancy sites are preferred over pits, revealing that O vacancies are the most interesting sites to elucidate the initiation and propagation of corrosion. 
To accurately compute ground and excited states of the O vacancy, we use the QDET approach and construct an effective Hamiltonian that we solve with full configuration interaction simulations on a classical computer and with the hybrid VQE algorithm and error mitigation on a quantum computer. 
When noise reduction is employed we find very good agreement with the FCI reference energy for ground and excited states. 

\section{\label{sec:CompDetail}Computational details}

\subsection{Structural relaxation}

While in many applications, Al$_2$O$_3$ occurs in its amorphous phase, for studies of defects and modeling hydration of surfaces, the corundum phase, $\alpha$-Al$_2$O$_3$, as the thermodynamically most stable polymorph is studied. 
For corundum, the Al-terminated (0001) surface is non-polar and is more stable than the O-terminated surface~\cite{GautierSoyer-1996,Guenard-1998} and is commonly employed in theoretical models to study the surface. 
Previous theoretical studies have concentrated on the pristine (0001) surface in a variety of environments consisting of both single atoms such as Al, O, Hf, Y, Pt, and S as well as molecule adsorption~\cite{Hinnemann-2007}, Cl ingress into the surface~\cite{Liu-2019}, carbon monoxide~\cite{Rohmann-2011} and hydrogen flouride~\cite{Quan-2012} adsorption, hydroxylated surfaces \cite{Chen-2022}, and adsorption of water~\cite{Ranea-2008,Thissen-2009,Wang-2011,Heiden-2019}. 
Few studies have concentrated on the near-surface vacancy in \alo{}~\cite{Liu-2019,Liu-2021,Chen-2023}, whereas hydration and vacancy interaction was studied in Ref.~\cite{Chen-2023}. 
To elucidate the initiation and propagation of corrosion in this material, we go beyond earlier works and present an in-depth analysis of not only the near-surface mono-vacancies and pits, but also the influence of hydration on these vacancies.

All the relaxations of atomic geometries and computation of the corresponding electronic properties were performed using density functional theory as implemented in the Vienna Ab-Initio Simulation Package (VASP)~\cite{Kresse-1993,Kresse-1996,Kresse-1996-2}.
We employ the projector augmented wave (PAW) method \cite{Blochl-1994,Kresse-1999} to describe the electron-ion interaction for Al (3$s^2$, 3$p^1$) and O (2$s^2$, 2$p^4$) valence electrons. 
We use PAW pseudopotentials with a higher energy cutoff, that are optimized for calculations with large numbers of unoccupied states and are suitable to compute excited-state properties using $GW$ simulations.
The structural relaxations were performed with the hybrid exchange-correlation functional HSE06~\cite{Krukau-2006}, using an energy cutoff of 600 eV for the plane-wave basis set.
To obtain the equilibrium lattice constant of the $\alpha$-Al$_2$O$_3$ structure in a 30-atom unit cell, we used a $\Gamma$-centered 6$\times$6$\times$2 $\mathbf{k}$-point mesh, and relaxed all forces on the atoms to less than 2 meV/\AA. 
The calculated bulk lattice constants of $a$=$b$=4.75 \AA\ and $c$=12.96 \AA\ are in very good agreement with experimental results of $a$=$b$=4.76 \AA\ and $c$=12.99 \AA\ \cite{Newnham-1962}.
Previous theoretical calculations with HSE06 using 32\,\% of exact exchange show similar accuracy, i.e.\ $a$=$b$=4.73 \AA, $c$=12.96 \AA\ \cite{Holder-2013} and $a$=$b$=4.74 \AA, $c$=12.95 \AA\ \cite{Gordon-2014}.

To create the \alo{} surface slab we repeat this bulk unit cell twice in the $x$ and $y$ direction, and introduce a vacuum of 25 \AA{} along the $z$ direction.
The resulting supercell consists of a total of 120 atoms with an Al terminated (0001) surface which is found to be the most stable termination~\cite{Guenard-1998}.
The atoms in the supercell are allowed to relax while the supercell dimensions were fixed at the equilibrium bulk lattice constant. 
To reduce the computational cost, the atomic relaxations in the supercell were performed until the forces on the atoms were less than 8 meV/\AA, employing a $\Gamma$-point sampling of the Brillouin zone.

\subsection{Quantum Defect Embedding Theory}

In this work, we follow the idea of quantum defect embedding theory (QDET) and divide the system into an active space Hamiltonian and environment. 
Within QDET the active space $A$ is defined such that the single-particle interactions of the defect states include the effects from the environment. 
The active space $A$ is selected encompassing the states localized near the defect, and is determined by calculating a localization factor \cite{Sheng-2022}
%-------------------------------------------------------------%
\setlength{\abovedisplayskip}{1pt}\setlength{\belowdisplayskip}{1pt}
\begin{equation}
 L(\psi_{n}^\mathrm{KS})= \int_{V\subseteq\Omega} |\psi_{n}^\mathrm{KS} (\textbf{x})|^{2}\,\mathrm{d\textbf{x}}.
\end{equation} 
%-------------------------------------------------------------%
Here, \textit{V} is a spherical region around the vacancy in the supercell with volume $\Omega$. 
The Kohn-Sham (KS) orbitals with a value $L$ larger than a specified threshold are selected as the active space. 
Orbitals not localized within $V$ are considered as the environment. 
We chose a sphere with $r$=1.4 \AA\ around the O vacancy to compute $L$. 
The largest active space studied in this work, composed of six occupied and six unoccupied bands, was obtained for a threshold $L$=10\,\%, and includes the minimum model with one occupied and one unoccupied defect state. 
Therefore, out of the total 576 bands, the largest active space has 12 bands and the minimum model has two bands.

The effective Hamiltonian for the active space, \Heff{}, is then defined as
\begin{equation}
H_{\mathrm{eff}}= \sum_{ij}^\mathrm{A} t_{ij}^{\mathrm{eff}}a_{i}^{\dagger}a_{j} + \frac{1}{2} \sum_{ijkl}^\mathrm{A} \nu_{ijkl}^{\mathrm{eff}}a_{i}^{\dagger}a_{j}^{\dagger}a_{l}a_{k}.
\label{Eq_Heff}
\end{equation}
Here, $a_i^{\dagger}$ and $a_j$ are creation and annihilation operators acting on the states in the active space $i,j,k,l$. 
The effective terms, $t_{\mathrm{eff}}$ and $\nu_{\mathrm{eff}}$, are renormalized one- and two-body integrals taking into account the screened Coulomb interaction between the active space and the environment. 
The dielectric screening from the environment is calculated within the random-phase approximation (RPA) \cite{Ma-Nature-2020,Nan-2022,Govoni-2018,Han-2019}.

To set up the Hamiltonian, ground-state simulations were performed with Quantum Espresso~\cite{Giannozzi-2009, Giannozzi-2017} using the plane-wave pseudopotential formalism and the Perdew-Burke-Ernzerhof (PBE) approach to describe exchange and correlation \cite{Perdew-1996}.
Norm-conserving pseudopotentials~\cite{Hamann-2013} for Al and O from the SG15 library~\cite{,Schlipf-2017} were used in these calculations. 
The kinetic energy cutoff was set to 90 Ry, and a total of 576 bands with $\Gamma$-point Brillouin zone sampling was employed for these DFT calculations.
Defect embedding simulations were carried out with the WEST code~\cite{Govoni-2015}, and a total of 954 eigenvectors were used to compute the static dielectric screening based on the projected-dielectric eigendecomposition (PDEP) of the dielectric matrix.
The PDEP method circumvents the need for the computationally costly summation over unoccupied states and also avoids the inversion of the dielectric matrix $\epsilon$ \cite{Wilson-2008}.

The resulting effective Hamiltonian \Heff{} is used to study the ground- and excited-state properties for a set of localized states which might host excitations of interest.
The ground- and excited-state properties for the active space are calculated by diagonalizing \Heff{} using full configuration interaction (FCI) simulations that provide an exact solution.
Full configuration interaction (FCI) calculations are performed on a classical computer for different sizes of the active space using PySCF~\cite{Sun-2018}.
Moreover, these FCI eigenvalues are employed as the reference energies for the subsequent analysis on a quantum computer.

To determine a \textit{minimum model}, we start by building an effective Hamiltonian for different sizes of the active space, and diagonalizing the Hamiltonian with full configuration interaction (FCI) to obtain the vertical excitation energies for the first two excited states.
In Fig.\ \supref{supfig:ActiveSpace} we plot the convergence of the FCI eigenvalues up to the third excitation energy for active spaces comprising of different numbers of occupied and unoccupied bands.
The \textit{minimum model} marked by the grey shaded area in Fig.~\supref{supfig:ActiveSpace} includes only two defect states (one occupied and one unoccupied state), corresponding to the states marked in Fig.~\figrefx{fig:Structure-HSE06}{b}.
From this, we find that the minimum model with two defect states is sufficient to describe the first two excited states. 
Larger active spaces always include the states from the minimum model. 
For the first and second excited state we see a convergence to within 5 and 20 meV, respectively (see Fig.\ \supref{supfig:ActiveSpace}).

The spin configurations of the first and second excited state of the effective Hamiltonian do not change with varying size of the active space (see Fig.\ \supref{supfig:ActiveSpace}).
These states are a triplet and a singlet state, respectively, whereas the third excitation is a triplet state except for the minimum model.
Additionally, we observe that including six unoccupied orbitals in the active space is sufficient to converge the energy of the third excited state, whereas the number of occupied states can be limited to only two.
This is due to the fact that all other occupied defect states are more than 60 eV below the occupied defect state in the minimum model and, hence, far from the band edge. 
On the other hand, the localized six unoccupied states lie in the energy interval of up to 3.5 eV from the first unoccupied defect state used in the minimum model. 
The minimum model is thus a good compromise between accuracy and computational cost, and allows for the subsequent quantum computer simulations.

\subsection{\label{subsec:QunatumSim}Quantum simulations}

%%------------------
\begin{figure*}
\label{fig:UCCAnsatz}
\includegraphics[width=0.95\textwidth]{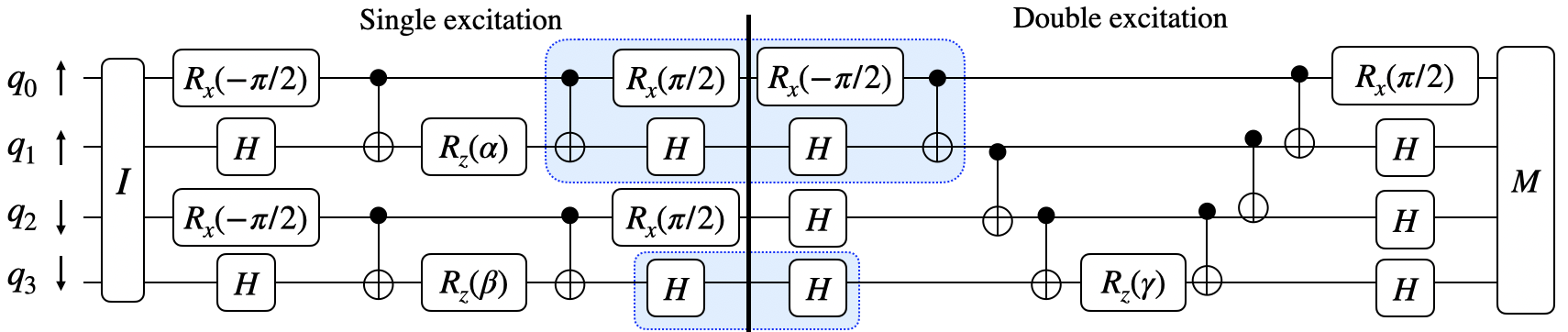}%
\caption{The UCC-3 ansatz \cite{McCaskey-2019} for a four-qubit circuit. 
The symbol $H$ represents a Hadamard gate. 
The $R_{x}$ and $R_{z}(\alpha,\beta,\gamma)$ represent rotation along the $x$ and $z$ direction, respectively, for the three variational parameters. 
The circuit initialization and measurement is denoted by box \textit{I} and \textit{M}, respectively. 
The four qubits $q_0,\dots,q_3$ with the respective spin channels ($\uparrow/\downarrow$) are also marked. 
Single and double excitations in the circuit are separated by the solid vertical line, and the blue shaded area highlights gates that will be identity gates and are not explicitly implemented after transpilation in the UCC-3 ansatz.
}

%\end{tikzpicture}
\end{figure*}
%%------------------

To benchmark noise characteristics of currently existing quantum processors we compute the ground- and excited-state energies of $H_\mathrm{eff}$, Eq.\ \eqref{Eq_Heff}, using a variational quantum eigensolver (VQE).
The VQE is a hybrid algorithm to calculate the lowest eigenvalue of a given Hamiltonian by employing a variational principle~\cite{Peruzzo-2014}.
Below we study the minimum model consisting of two electrons, i.e.\ four spin orbitals, on a four-qubit quantum circuit. 
The qubit operator for the Hamiltonian is mapped by the Jordon-Wigner transformation~\cite{Jordan1934} and expressed in terms of Pauli matrices $\{I, X, Y, Z \}$.
The wave function was prepared by a parameterized Hartree-Fock state using the unitary coupled-cluster, UCC-3 ansatz~\cite{McCaskey-2019}, as shown in Fig.~\ref{fig:UCCAnsatz}.

The circuit in Fig.\ \ref{fig:UCCAnsatz} is initialized in block $I$ with $\vert 1010\rangle$ as the reference Hartree-Fock ground state configuration~\cite{McCaskey-2019}. 
In the UCC-3 ansatz, single excitations from the ground state are represented by the first half of the circuit, parameterized by the $R_{z}(\alpha)$ and $R_{z}(\beta)$ gates.
Double excitations are described by the second half of the circuit, parameterized by the $R_{z}(\gamma)$ gate. 
This ensures that all possible excitations of the two electrons in the ansatz are accounted for in the Hamiltonian~\cite{McCaskey-2019}, hence, this ansatz is a complete variational basis for this two-electron problem. 
The energy
\begin{equation}
\label{eq:energy}
E(\theta)= \langle \psi_{0} \vert H_\textrm{eff} \vert \psi_{0} \rangle
\end{equation}
serves as a cost function when updating the parameters classically with the COBYLA~\cite{Powell-1994} optimizer, where $\theta$ refers to the parameters in the circuit $\{\alpha,\beta,\gamma\}$.

To find excited-state eigenvalues, the weighted subspace-search variational quantum eigensolver (SSVQE) \cite{nakanishi2019subspace} is used on quantum processors.
Since the UCC-3 ansatz is designed for ground state calculations, we instead use the UCC-1 ansatz from Ref.~\cite{McCaskey-2019},  $U(\theta)$, which corresponds to the second half of the UCC-3 ansatz and only contains the double excitations in Fig.\ \ref{fig:UCCAnsatz}, to parameterize the wave function. 
In SSVQE, excited-state properties are evaluated by imposing the orthogonality between the UCC-1 ground state and excited states.
For measuring the energy of the $k$-th excited state, $k$ different orthogonal initial states  $\{{|\psi_j\rangle}\}_{j=1}^k$  are required, and SSVQE is realized by minimizing the cost function
%%------------------
\begin{equation}
\label{ssvqe_cost}
L(\theta)= \sum_{j=0}^k w_j \langle \psi_j |U(\theta)^{\dagger}HU(\theta)| \psi_j \rangle,
\end{equation}
%%------------------
where $w_j$ is the weighting factor for the expectation value.
The energy eigenvalues do not depend on the specific choice of $w_j$, as long as $w_i < w_j$ for $i>j$, as discussed in Ref.\ \cite{nakanishi2019subspace}.

The Qiskit~\cite{Qiskit-2024} Runtime Estimator primitive service is used with the \textit{session} tag, which allows an uninterrupted iterative communication with the hardware during execution of the hybrid quantum-classical VQE simulations.
The Qiskit Runtime Estimator measures the ansatz circuit 10,000 times and outputs one single energy expectation value, Eq.\ \eqref{eq:energy}, from these. 
Readout error mitigation and no transpilation was employed for the quantum circuit.
The noise model from the hardware is employed with the QASM and the AER simulators of the Qiskit framework~\cite{Qiskit-2024} to calculate the ground-state energy with error mitigation. 
For the excited-state properties, the actual quantum simulations are performed on the ibm\_osaka hardware, comprising of 127 qubits.

\subsection{\label{subsec:ErrorMitigation}Error mitigation}
In this work we use zero noise extrapolation (ZNE) \cite{Li-PRX-2017} to mitigate errors introduced by noise that occurs on real quantum hardware. 
The error mitigation process within ZNE involves systematically scaling up the noise in a circuit by inserting identities and extrapolating measured results for observables to obtain the zero-noise limit. 
In this work, scaling up the noise is achieved by the global folding technique, and the corresponding circuit is illustrated in Fig.~\supref{supfig:ZNE}.
For simplicity, Fig.\ \supref{supfig:ZNE} illustrates ZNE for two qubits, whereas in this work, $U$ represents the four qubit UCC-3 ansatz.

In Fig.\ \supref{supfig:ZNE}, the blue colored box represents one $UU^{\dagger}$ operator, which corresponds to the identity operator in the noiseless regime.
On real quantum hardware, however, each identity introduces noise and we scale up the noise by repeating the identity block. 
We then analyze the error in the ground-state energy, calculated as the difference of the VQE and FCI energy, with respect to the number of identities.
In our circuit, two-qubit CNOT gates are considered as the main source of noise, while  contributions from the other gates in the ansatz are assumed to be negligible~\cite{Li-PRX-2017}.
In the UCC-3 ansatz we have eight CNOT gates (see Fig.~\ref{fig:UCCAnsatz}), and the energy with increased noise is plotted as a function of number of CNOT gates, calculated as $8\times(2n+1)$.
Here, $n=0$ is the case with no global folding, and $n=1,2,\dots$ is the number of  identity blocks added to boost the noise. 

Besides VQE, we also use a post-selection technique to mitigate the error in the results for both ground- and excited-state calculations.
Post selection relies on the conservation of number of particles in a system, which is violated in the presence of noise.
For our studied case, the number of electrons is fixed, and the measured results can be post-processed based on this constraint.

However, this principle holds true only when performing measurements in the $Z$ basis, which is the standard computational basis \cite{mermin2007quantum}, as the measurement of the $X$ and $Y$ basis does not commute with the particle number operator. 
In order to measure all qubit operators in the $Z$ basis and to apply the post-selection technique, we convert the $X_pX_q+Y_pY_q$ qubit operator, originating from the off-diagonal elements of the Hamiltonian, into the $I_pZ_q-Z_pI_q$ operator by implementing a unitary quantum gate as described in Ref.\ \onlinecite{google2020hartree},
%%------------------
\begin{figure}[!htp]
\includegraphics[width=0.95\columnwidth]{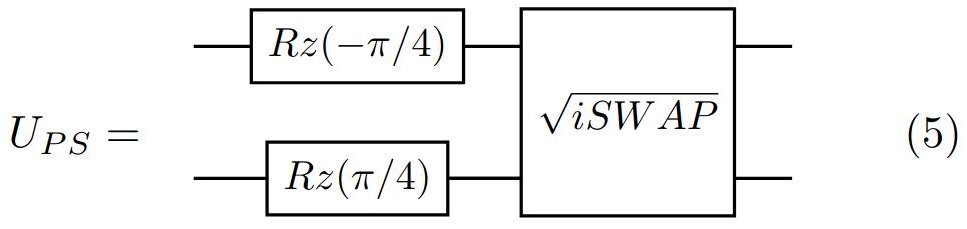}%
\end{figure}
%-------------------------------------------------------------%
%%------------------
This circuit is concatenated with our ansatz and subsequently we run the VQE simulations.

\section{\label{sec:Properties-Surface}Structural properties}

\subsection{Pristine surface}

First, we validate our results for relaxed atomic geometries and surface energies for the \alo{} pristine surface by comparing to prior literature data.
Formation of the surface results in 85\,\% inwards relaxation of the surface Al layer when compared to the bulk positions (see Table \ref{tab:Displacement}).
In our calculation, we observe a reduction from 1.85 \AA{} to 1.68 \AA{} concurrent with the trend reported in Ref.\ \onlinecite{Rohmann-2011}.
Compared to the experimental value of 51\,\% reported in Ref.\ \onlinecite{Guenard-1998}, an overestimation of the Al-O distance comparable to ours has been observed in previous theoretical calculations, reporting 90\,\% \cite{Carrasco-2004} and 93.8\,\% \cite{Rohmann-2011}.
This large inward relaxation of the surface Al layer was attributed to the charge redistribution from Al to O by Gautier-Soyer \etal{}~\cite{GautierSoyer-1996} and in theoretical calculations, it is manifested as a reduction of the surface Al$-$O bond length.
We note that the effect of the environment discussed in Ref.~\cite{Wang-2000} indicates that saturating the surface by placing an H atom on top of the surface O atom reduces the relaxation of the surface Al-O distance from 86 \% to  $\sim$69 \%. 
The inter-planar spacings of Al-O and O-O for the sub-surface layers in Table~\ref{tab:Displacement} show good agreement with the trends observed in previous theoretical calculations~\cite{Carrasco-2004,Rohmann-2011} and experiments~\cite{Guenard-1998}, wherein an alternating compression and expansion is observed from the surface to the sub-surface layers.

The calculated surface energy for the pristine surface is also consistent with previous theoretical works.
We obtained a value of 1.69 J/m$^2$ in our simulations, compared to the experimental value of 2.64 J/m$^2$~\cite{McHale-1997}.
Previously reported theoretical surface energies of 1.55 J/m$^2$~\cite{Carrasco-2004}, 1.9 J/m$^2$~\cite{Sun-2006}, and 1.51 J/m$^2$~\cite{Rohmann-2011} also yield slight underestimates.
Our results agree well with these other simulations, but differences relative to experiment might be attributed to the dissimilar surface termination at the top and bottom of our slab, and the effect of surface contamination in experiment~\cite{Wang-2000}.
Overall, our final geometry is in agreement with the literature and we use this to study surface defects going forward.

%-------------------------------------------------------------%
\begin{table}[ht]
\caption{\label{tab:Displacement}
Effect of surface relaxation: 
Change in interplanar spacing, ($d_{\mathrm{relaxed}}-d_{\mathrm{bulk}}$)/($d_{\mathrm{bulk}}$), in \% along the [001] direction upon relaxation.
%The unrelaxed distances correspond to bulk distances.
A negative sign indicates a decrease of the interplanar distance relative to bulk.
%whilst a positive sign indicates an outward relaxation.
The layer numbering is indicated by the subscript starting from the top surface.
The entire slab consists of 12 Al layers.
}
\begin{ruledtabular}
\begin{tabular}{lcccc}
Layers & This work &  \multicolumn{2}{c}{Previous work}  & Exp. \\[0.5ex]
 & & Ref.\ \cite{Rohmann-2011} & Ref.\ \cite{Carrasco-2004} & Ref.\ \cite{Guenard-1998}\\[0.5ex]
\colrule
%-------------------------------------------------------------%
Al$_1$/O$_2$ & $-85.0$ & $-90.0$ & $-93.8$ & $-51$ \\[0.5ex]
O$_2$/Al$_3$ & $+3.5$ & $+3.19$ & $+6.1$ & $+16$ \\[0.5ex]
Al$_3$/Al$_4$ & $-45.8$ & $-45.1$ & $-46.7$ & $-29$ \\[0.5ex]
Al$_4$/O$_5$ & $+20.2$ & $+20.2$ & $+22.0$ & $+20$ \\[0.5ex]
 %-------------------------------------------------------------%
\end{tabular}
\end{ruledtabular}
\end{table}
 %-------------------------------------------------------------%

\subsection{\label{sec:surfacedefects}Surface defects: Mono-vacancies and pit}

Hydration is often considered as a precursor to corrosion and in this work we study the role of vacancies and hydration of the \alo{} surface in corrosion of aluminum. 
We first compare ground-state total energies and formation energies of different mono-vacancies that are potential candidates for initial states of pit formation.
Through removal of surface atoms we focus on Al and O vacancies near the surface. 
The two starting configurations include a supercell with a near-surface O vacancy ($V_{\mathrm{O}}$) and another configuration with an Al vacancy ($V_{\mathrm{Al}}$) on the surface. 
Subsequent pit formation is represented by removal of an additional Al atom from the subsurface layer of the energetically most favorable mono-vacancy.
Hydration is represented in our simulations using an O adatom near the surface, to reduce the complexity of modeling a water molecule.
A single water molecule may be represented as an O atom to accommodate the three-body breakup process of water~\cite{Haxton-2008}, e.g.\ at higher altitudes due to the strong UV radiation.
For the pit configuration as well as all mono-vacancies, we also compute formation energy differences between hydrated and non-hydrated state.

We validate our atomic geometries of these defect configurations and find good agreement with prior DFT-PBE simulations:
On the one hand, introducing an O vacancy on the surface results in an inward relaxation of the adjacent surface Al atom by 14\,\%.
On the other hand, introducing an Al vacancy leads to the outward relaxation of nearest surface O atoms by $\approx 1.5$\,\%. 
These results are in close agreement with 13.3 \% outward and 2.2 \% inward relaxation, respectively, from the work of Carrasco \etal{}~\cite{Carrasco-2004}

The calculated $V_{\mathrm{O}}$ formation energy  is 8.48 eV, which is $\sim$6 eV lower than that of $V_{\mathrm{Al}}$, and within 0.05\,\% and 0.01\,\%, respectively, of the values reported in Ref.\ \onlinecite{Carrasco-2004}.
A smaller formation energy of the $V_{\mathrm{O}}$ suggests that pits likely start to form around O vacancies, because they would be more prevalent than Al vacancies.
Therefore, to simulate pit formation that occurs during corrosion, we started from $V_{\mathrm{O}}$ and additionally removed an Al atom from the sub-surface layer to represent the pit, denoted as $V_{\mathrm{O+Al}}$.
Upon relaxation of the pit, the maximum local distortion involves a vertical inward relaxation by 0.9 \AA{} of the surface Al atom nearest to the surface O vacancy.
We use these geometries in the following as starting point for a hydrated surface that represents aqueous environments.

%-------------------------------------------------------------% 
\begin{figure}
\includegraphics[width=0.95\columnwidth]{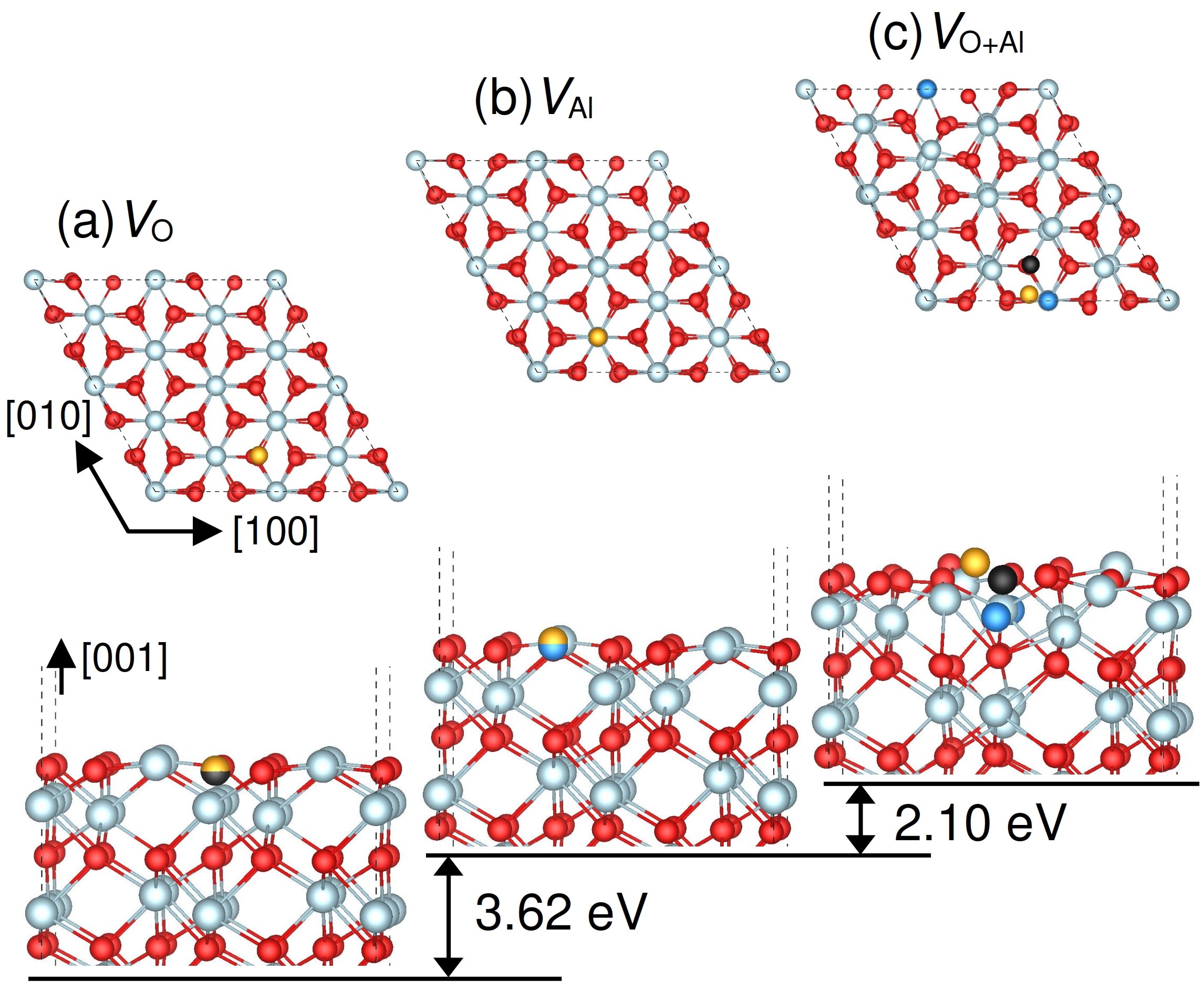}%
\caption{\label{fig:OAdatom-HSE06}
Difference of the total energy for $V_\mathrm{O}$, $V_\mathrm{Al}$, and $V_\mathrm{O+Al}$ relaxed geometries, and the respective relaxed geometry after adding an O adatom.
The pristine surface has the same total energy as adatom+$V_\mathrm{O}$ after relaxation and is used as energy zero.
Top and side view of the relaxed $\alpha$-Al$_2$O$_3$ (0001) surface geometries for $V_\mathrm{O}$ (a), $V_\mathrm{Al}$ (b), and $V_\mathrm{O+Al}$ (c).
Al and O atom are marked by light blue and red spheres, Al and O vacancies are marked by dark blue and black spheres, respectively.
The O adatom is shown in orange.
%\as{Remove axes.}
%\as{Remove dashed lines in bottom figure.}
%\as{Move direction labels to the left of the figure.}
%\as{Move top figure further apart from bottom.}
%\vbh{Done}
}
\end{figure}
%-------------------------------------------------------------%

After relaxing hydrated defect configurations, with an additional O adatom on top of the vacancies, we calculated the adsorption energy as the total energy difference between the relaxed configurations with and without the O adatom,
\begin{equation}
    E_\mathrm{ads} = E_{\textrm{tot(Defect+O)}}-E_{\textrm{tot(Defect)}}.
\end{equation}
For the surface with an oxygen vacancy, the energy difference due to hydration is largest among the defects studied here and the relaxed configuration recreates the ideal surface. 
A recent study~\cite{Chen-2023} also shows that OH$^-$ relaxes into the O vacancy on the surface and supports our simple picture of employing an O adatom to represent hydration. 
Previous studies also demonstrated that an adsorbed Cl atom~\cite{Liu-2019,Liu-2021} prefers to relax into a surface O vacancy, rather than a surface Al vacancy. 
Since the total energy of the relaxed $V_\mathrm{O}$ configuration with the adatom is the same as the pristine surface, it is used as energy zero for subsequent analysis [see Fig.~\figrefx{fig:OAdatom-HSE06}{a}]. 

For the $V_{\mathrm{Al}}$ configuration [see Fig.~\figrefx{fig:OAdatom-HSE06}{b}], the O adatom relaxes into the $V_{\mathrm{Al}}$ position, and the adsorption energy is 3.62 eV larger than for the $V_{\mathrm{O}}$ configuration. 
For the $V_{\mathrm{O+Al}}$ configuration [see Fig.~\figrefx{fig:OAdatom-HSE06}{c}], the O adatom prefers to move towards the $V_\mathrm{Al}$ in the subsurface layer but never relaxing into the pit.
At the same time the Al surface atom adjacent to the $V_\mathrm{O}$ undergoes an outward vertical relaxation by 0.3 \AA{}.
The adsorption energy is 5.72 eV larger, compared to $V_\mathrm{O}$. 

We hypothesize that due to the lower formation energy of the O mono vacancy, they are more likely to form on the surface. % and start pitting due to various external factors. 
Since the relaxation of the O adatom into the O vacancy restores the pristine surface, presumably only a fraction will start pitting during hydration. 
These pits persist even after hydration, indicating that corrosion progresses in these spots and into the material. 
Therefore, we consider the O vacancies to be the primary defect sites for the initiation of corrosion. 
For this reason, an in-depth analysis of the near-surface $V_{\mathrm{O}}$ is presented in the remainder of the work.
We study the ground- and excited-state properties of this defect configuration with classical and quantum computers.

%-------------------------------------------------------------%
\begin{figure}
\begin{center}
\includegraphics[width=0.5\textwidth]{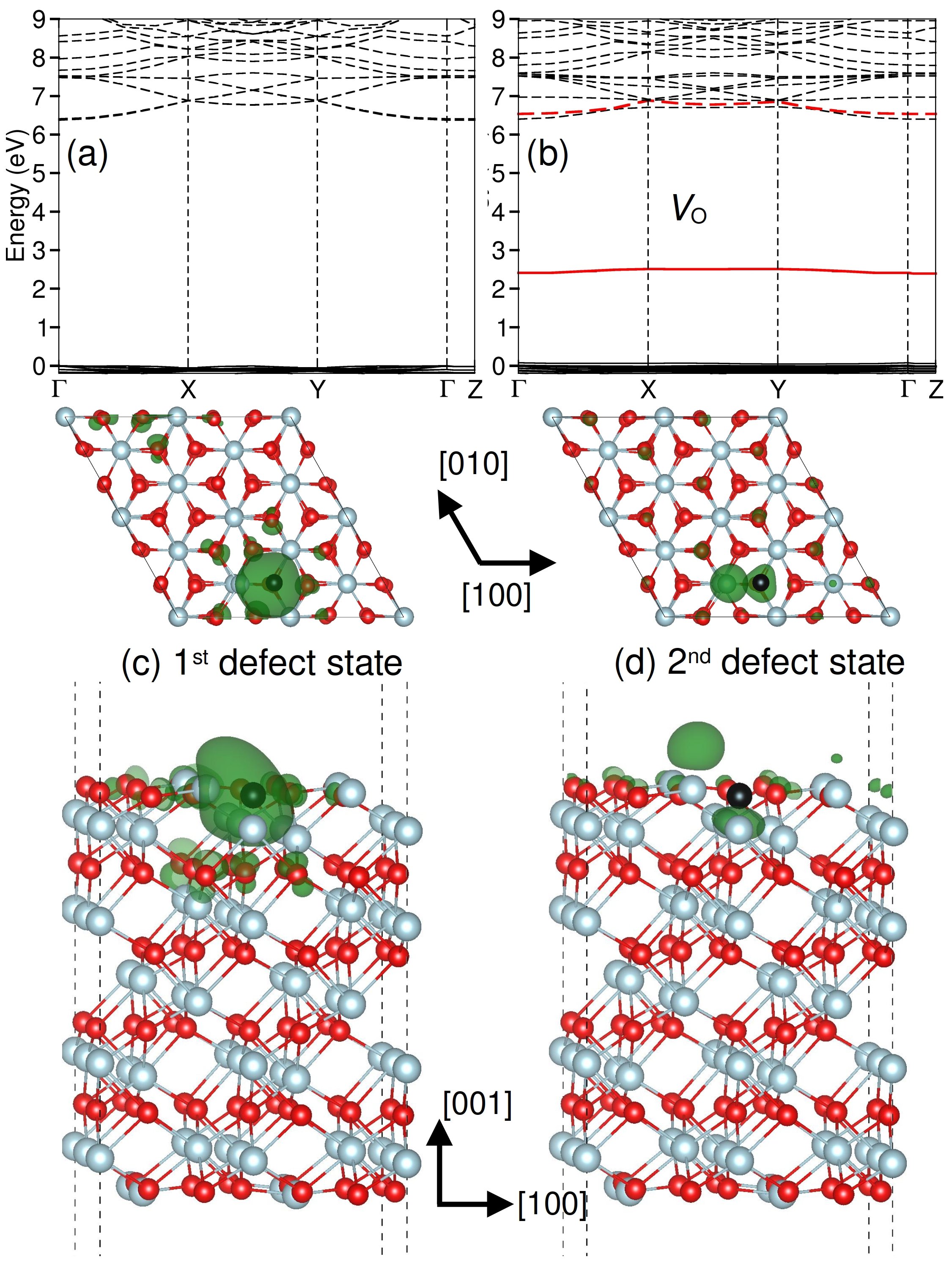}%
\caption{\label{fig:Structure-HSE06}
Band structure of the (a) pristine surface and (b) the surface with an O vacancy, respectively, where the valence band maximum is set to zero.
Occupied and unoccupied states are marked in solid and dashed lines, respectively. 
For the O vacancy, defect states are marked in red.
Band-decomposed charge density for the occupied defect state (c), and the first unoccupied defect state (d) marked in (b).
}
\end{center}
\end{figure}
%-------------------------------------------------------------% 

\section{\label{sec:Properties-Surface}Electronic properties of a surface oxygen vacancy}

Introducing an oxygen vacancy leads to a defect state inside the band gap of 6.38 eV of the pristine surface (see Fig.\ \ref{fig:Structure-HSE06}).
This gap is slightly smaller than the experimental value of 6.7 eV~\cite{Andersson-1999} and remains unchanged upon introduction of the O vacancy.
The band dispersion of the conduction states also remains mostly unaltered compared to the pristine surface, except for the bands in the energy range of 6.2 to 8 eV, as shown in Fig.~\figrefxy{fig:Structure-HSE06}{a}{b}.
The changes are small for the states above 7.5 eV, because these bands originate from largely undistorted Al atoms far from the vacancy.
In addition, states at 7 and 7.5 eV are affected more as they correspond to the Al atom near the vacancy that experiences inward relaxation (see Sec.\ \ref{sec:surfacedefects}).

Due to the presence of the $V_{\mathrm{O}}$, defect states emerge in the band structure at $\sim$2.45 eV (occupied state) and 6.53 eV (unoccupied state), marked in red in Fig.\ \figrefx{fig:Structure-HSE06}{b}.
These defect states are characterized by a distinct non-dispersive behavior in reciprocal space, and are almost flat around the $\Gamma$ point. 
Also, the band decomposed charge density for the two defect states is plotted in Fig.~\figrefxy{fig:Structure-HSE06}{c}{d}.
A strong localization of the charge density is observed near the O vacancy and the adjacent surface Al for both defect states marked in Fig.~\figrefx{fig:Structure-HSE06}{b}.

Next, we report more accurate vertical excitation energies, including many-body effects, from QDET calculations for the $V_{\mathrm{O}}$ configuration.
These defect states are confined only to a small space, hence quantum embedding approaches are suitable to accurately compute these strongly localized states at a high level of theory. 
In the remainder of the paper, we compute the first and second excitation energy employing the QDET approach for the $V_{\mathrm{O}}$ configuration, and report the corresponding vertical excitation energies. 

\section{\label{sec:QuantumSim} Quantum simulations on a quantum computer}

\subsection{\label{subsec:GS on QC} Ground-state energy}

%--------------------------------------
\begin{figure}
\begin{center}
\includegraphics[width=0.49\textwidth]{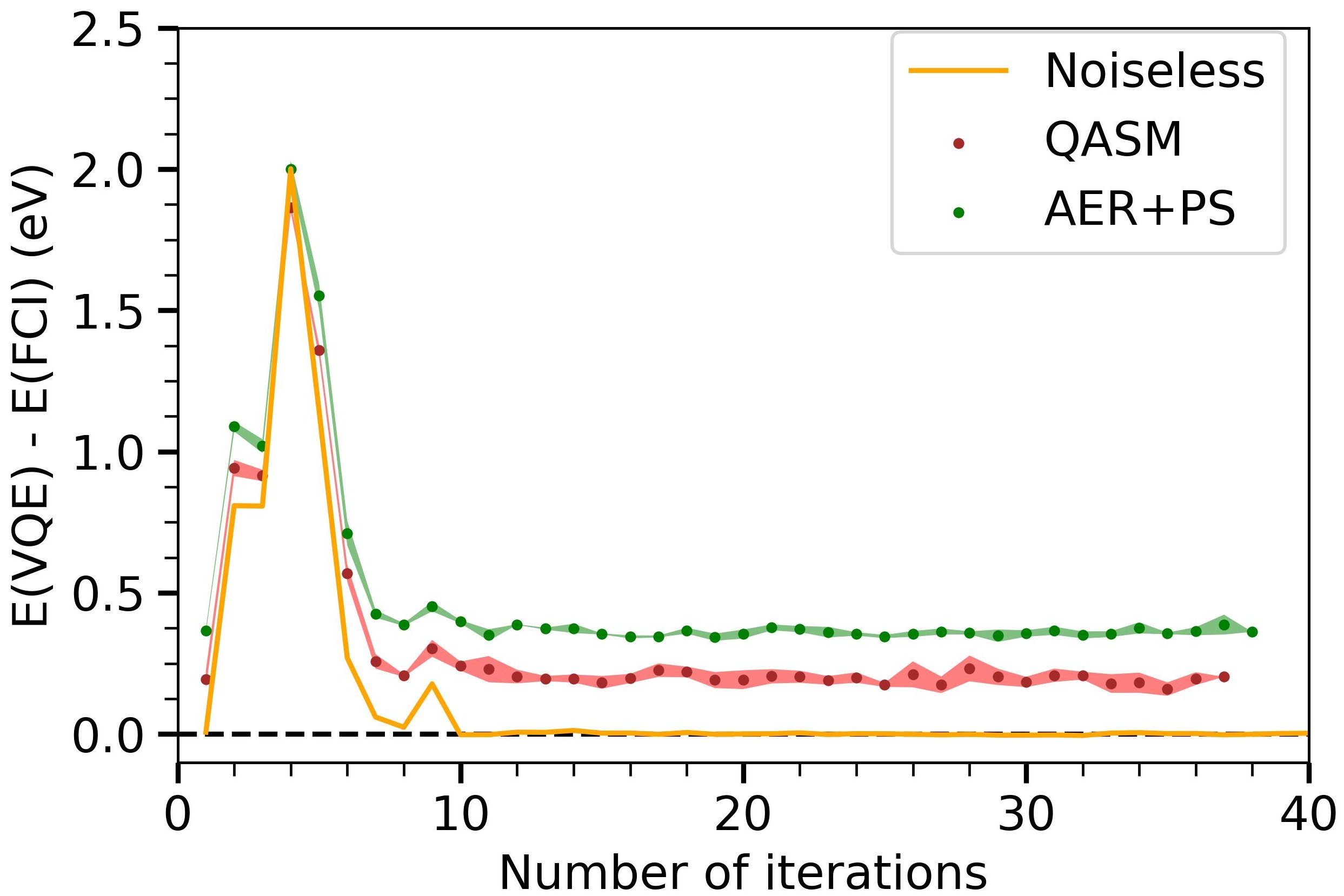}%
\caption{\label{fig:QC-Groundstate}
Convergence of the ground-state energy difference between the VQE and FCI values for the minimum model with four qubits employing VQE on the QASM and AER simulators, with the noise model from ibm\_osaka.
The energy from the noiseless QASM simulator is plotted as orange solid line.
Agreement between FCI and VQE is drawn as black dashed line.
The shaded red and green areas show the standard deviation of three independent calculations with QASM and AER, respectively.
The respective mean energies are plotted with red and green filled circles. 
}
\end{center}
\end{figure}
%--------------------------------------

To validate our implementation, we compare the calculated ground-state energy with a noiseless simulator to full configuration interaction (FCI) reference values and find excellent agreement.
The minimum model active space, comprising of two electrons for the $V_{\mathrm{O}}$ configuration, is used (see Sec.\ \ref{sec:CompDetail}).
The ground state is calculated on a simulated four-qubit circuit using the UCC-3 ansatz. 
In Fig.~\ref{fig:QC-Groundstate}, we plot the ground-state energy difference between the VQE and FCI results with respect to the number of VQE iteration.
With the noiseless QASM simulator of Qiskit, we obtain excellent agreement with the reference energy, and the energy converges already at the tenth iteration.
The good agreement with FCI also shows that the UCC-3 ansatz employed in this work successfully evaluates the exact ground state, as expected for two electrons.

Next we incorporate a noise model and find that the ground-state energy does not converge to the FCI value.
We use the noise model of the 127-qubit ibm\_osaka hardware, and calculate the ground-state energy with the QASM and AER simulators in Qiskit.
Three independent calculations were performed on both simulators, and we plot the standard deviation (shaded area) and mean (filled circles) of these runs in Fig.\ \ref{fig:QC-Groundstate}.
We note that not all VQE simulations have the same number of iterations, and hence the standard deviations are absent towards the end of the respective simulator plots.  
For the AER simulations, post-selection was necessary to conserve the number of electrons and obtain energies close to FCI.
The QASM and AER simulators render energies with an error of 190 and 375 meV, which is about five and nine times larger than the chemical accuracy of 40 meV, respectively. 
Such errors are predominantly attributed to the two-qubit CNOT gates \cite{Li-PRX-2017} and we discuss zero-noise extrapolation (ZNE) in the next section as a successful error-mitigation technique. 

\subsection{ZNE error mitigation}
%-------------------------------------------------------------%
\begin{figure}
\includegraphics[width=0.95\columnwidth]{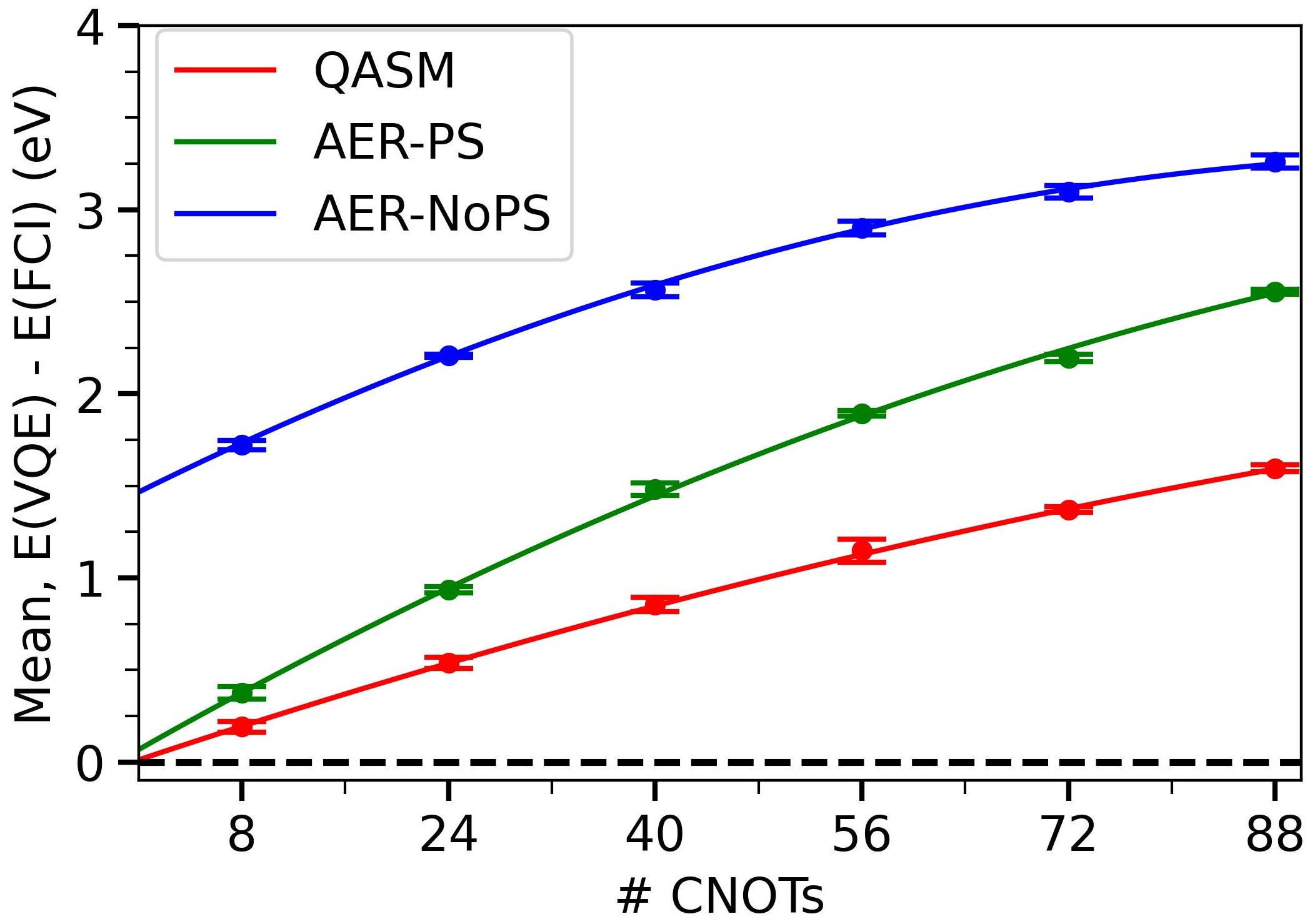}%
\caption{\label{fig:QC-ZNE}
ZNE error mitigation of the ground-state energy.
The QASM (blue color) and AER simulator were used with noise models from ibm\_osaka.
The mean error in energy of three runs is shown as solid circles along with the respective standard deviation.
The AER simulator was tested with (green color) and without (blue color) a post-selection (PS) approach.
The second-order polynomial fit extrapolated to the zero-CNOT case is represented by the solid lines.
Agreement between FCI and VQE is drawn as black dashed line.
}
\end{figure}
%-------------------------------------------------------------%

Employing the ZNE approach described in Sec.~\ref{subsec:ErrorMitigation} we mitigate the error for the QASM and AER simulators.
The error in energy is plotted as a function of number of CNOT gates in Fig.\ \ref{fig:QC-ZNE} and is fitted to a second-degree polynomial and extrapolated to the zero-CNOT case, \ie{} the zero-noise limit. 
In Fig.~\ref{fig:QC-ZNE}, we observe a near-linear scaling for the circuit with up to 40 CNOT gates.
We find that using ZNE brings the ground-state energy from the QASM simulator within the chemical accuracy of the FCI reference.
%of 40 meV 
With the QASM simulator, the error reduces to 11$\pm 36$ meV, whereas the error without ZNE is 191$\pm 29$ meV for the original circuit.  
On the AER simulator with post selection, the error after correction drops to 66$\pm 35$ meV from 376$\pm 34$ meV.
However, without post-selection, the error remains almost uncorrected to 1465.4$\pm 34$ meV, from 1721.8$\pm 25$ meV.
We note that with a large number of CNOT gates ($>$40) the errors approach a plateau where the noise of the CNOT gate is significant, leading to saturation and a barren plateau that is typically observed at very high noise~\cite{Li-PRX-2017}. 
In summary, our results confirm that the straightforward ZNE approach can yield ground-state energy within chemical accuracy, without the need of additional qubits \cite{Huggins-2021}.
Although VQE suffers from the requirement of large number of measurements to accurately estimate the energy, it has the advantage that it can execute a large number of independent, short quantum simulations, and is more suitable for current NISQ computers than the quantum phase estimation technique which requires long coherent circuits~\cite{Bauer-2020}.

\section{Excited-state properties from subspace-search VQE}

%-------------------------------------------------------------%
\begin{figure}
\includegraphics[width=0.95\columnwidth]{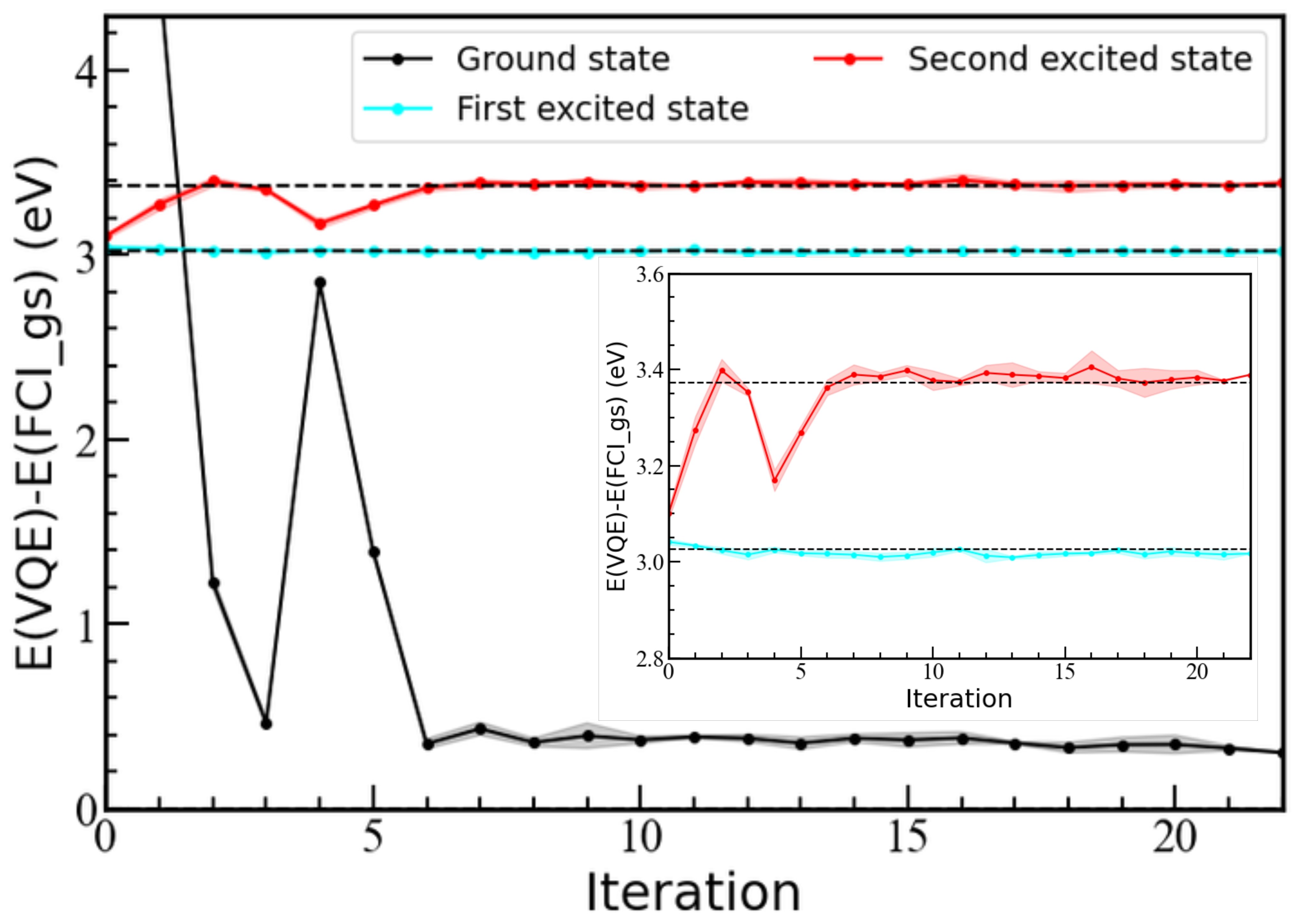}%
\caption{\label{fig:SSVQE}
Calculation of the first and second excited state with subspace-search VQE and post selection on ibm\_osaka.
The average of four runs (solid circles) and the respective standard deviation (shaded area) of the ground state, and first and second excited state are marked in black, cyan, and red, respectively. 
Reference FCI energies for the respective states are marked by a black dashed line.
The inset magnifies around excited states.
}
\end{figure}
%-------------------------------------------------------------%

Next, we investigate the first and second excited-state energies, computed using subspace-search VQE (SSVQE) and post-selection error mitigation, and find these to be within chemical accuracy of the FCI eigenenergies.
Here, for the weights $w$ in the cost function of SSVQE in Eq.\ \eqref{ssvqe_cost}, we used 1, 0.8, and 0.5, for the ground state, first excited state, and second excited state, respectively.
For the first and second excited state, the converged energies are 3.020$\pm$0.007 eV and 3.390$\pm$0.013 eV, respectively.
A previous work on the $NV^{-}$ center in diamond and the $VV$ configuration in 4\textit{H} SiC~\cite{Huang-2022} also successfully calculated the excited state energies using a similar quantum subspace expansion.
Notably, the ground state energy within this approach shows an error of around 350$\pm$0.040 meV, and is in poor agreement with the VQE+ZNE calculations discussed in the previous section.
The large discrepancy is attributed to the use of the UCC-1 ansatz, which provides an insufficient description of the ground state compared to the UCC-3 ansatz due to the absence of single-excitation terms. 
However, this ansatz is sufficient for describing the first two excited states, as they already involve the excitation of one electron when constructing the orthogonal basis states in Eq.\ \eqref{ssvqe_cost}.
Our results confirm the utility of using this algorithm to compute excited-state energies. Notably, the algorithm's effectiveness is independent of ground-state results, provided the description of the excited states is sufficiently accurate.

\section{\label{sec:summary}Conclusions}

%CHECK::\supref{supfig:ActiveSpace}
We present computational results for the ground- and excited-state electronic properties of O vacancy defects near the (0001) surface of \alo{}, computed with the full configuration interaction technique as well as quantum defect embedding theory.
We compare the pristine surface to one in aqueous environment and find that although an O vacancy may be repaired during hydration, pitting initiated in these sites remains exposed and likely causes further degradation of the material.
Using a variational quantum eigensolver, in addition to the full configuration interaction approach, we calculate the electronic properties of the $V_{\mathrm{O}}$ configuration, and report the first and second excited states. 
Interestingly, in this work for the O vacancy on the surface, the first excited state is in the visible range and is a triplet state indicating a long-lived excitation that can be exploited as a platform for quantum information science applications and for photocatalytic degradation.

On a noiseless simulator, for the minimum model with two electrons we report excellent agreement of the ground-state energy with the FCI reference energy. 
Our simulations show that zero-noise extrapolation error mitigation of the QASM simulator results reduces the noise due to the use of quantum hardware for the ground state properties, and agreement with the exact solution is achieved within chemical accuracy.
Moreover, by employing a post selection approach with the weighted subspace-search VQE allows for the correct evaluation of the first and second excited state of the minimum model. 
Our work extends the use of quantum defect embedding theory and quantum computing hardware from bulk to surfaces and we believe that our results serve as a benchmark to describe electronic properties with quantum computers for other surfaces with defects.

\begin{acknowledgments}
We acknowledge funding by the IBM-Illinois Discovery Accelerator Institute. 
The calculations were performed on the Illinois Campus Cluster, a computing resource that is operated by the Illinois Campus Cluster Program (ICCP) in conjunction with the National Center for Supercomputing Applications (NCSA) and which is supported by funds from the University of Illinois at Urbana-Champaign.
\end{acknowledgments}

%\bibliography{aipsamp}
\providecommand{\noopsort}[1]{}\providecommand{\singleletter}[1]{#1}%

\clearpage{}
%-------------------------------------------------------------%

\beginsupplement

\section*{Supplement}

\subsection{Convergence of active space}
We plot the convergence for the active space in the QDET calculations. 
%-------------------------------------------------------------%
\begin{Suppfig}
\label{supfig:ActiveSpace}
\begin{figure}[!htp]
\begin{center}
\includegraphics[width=0.43\textwidth]{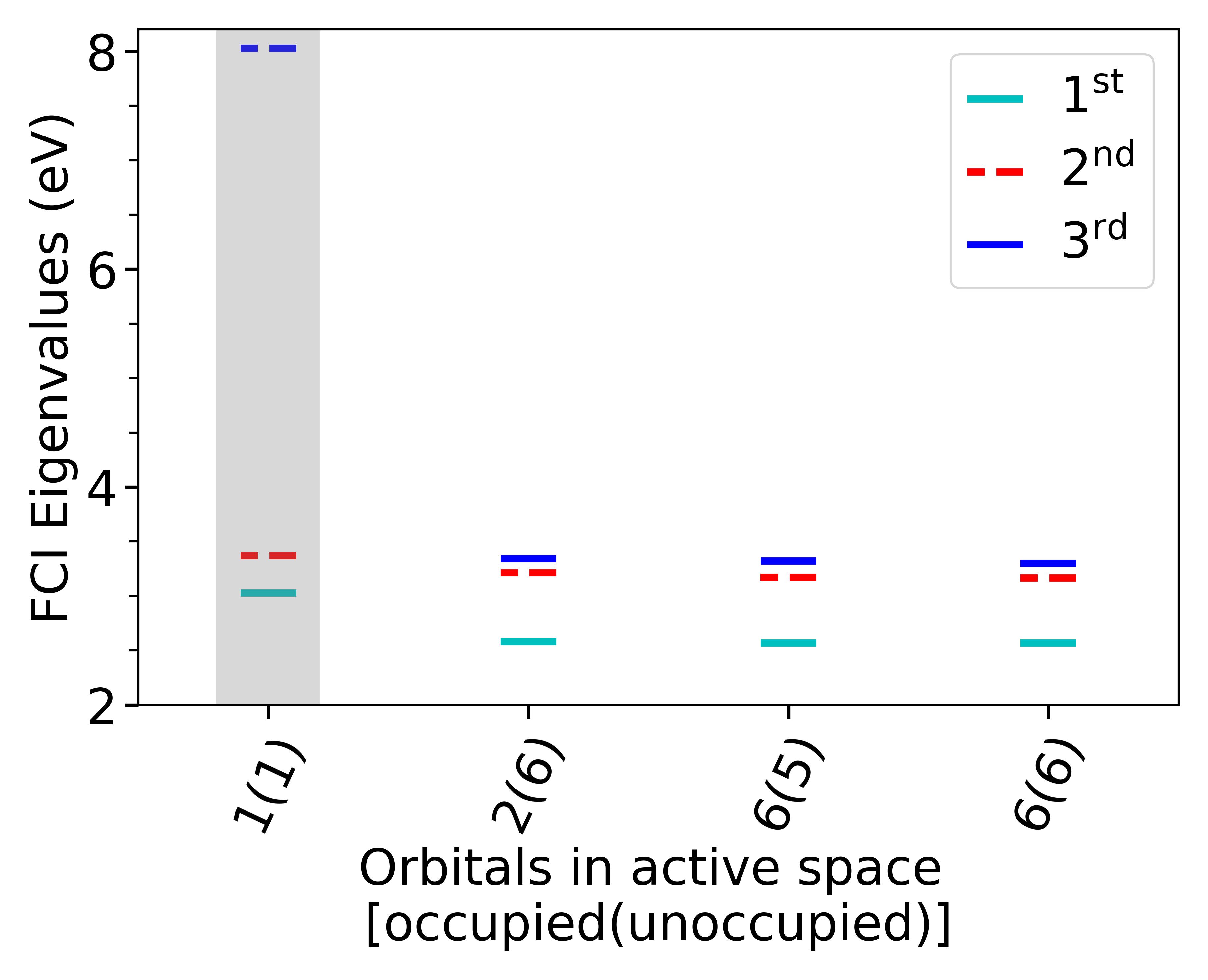}%
\caption{
Convergence of excitation energies obtained from Full configuration interaction (FCI) calculations with respect to the size of the active space comprising of combinations of occupied(unoccupied) bands determined from quantum embedding.
The first, second and third excited state are shown in cyan, red and green respectively. The triplet and singlet excited state are marked by a solid and dashed line, respectively. The gray shaded area marks the  minimum model with one occupied and one unoccupied band.
}
\end{center}
\end{figure}

\end{Suppfig}

\subsection{Global folding for zero-noise extrapolation}
\begin{Suppfig}
\label{supfig:ZNE}
%-------------------------------------------------------------%
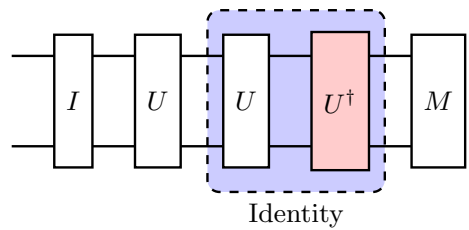
\begin{figure}[!htp]
\begin{center}
\begin{adjustbox}{width=0.35\textwidth}
\begin{quantikz}
& \gate[2]{I} & \gate[2]{U}&
\gate[2]{U}\gategroup[2,steps=2,style={dashed,rounded
corners,fill=blue!20, inner
xsep=2pt},background,label style={label
position=below,anchor=north,yshift=-0.2cm}]{Identity} &
\gate[style={fill=red!20},2]{U^{\dagger}}  & \gate[2]{M}\\
& & &  &  &
\end{quantikz}
\end{adjustbox}
\caption{Illustration of zero-noise extrapolation for two qubits, where $U$ is a circuit, and the identity block is shown in the blue block.
Each identity block introduces noise systematically that is used for extrapolation to the zero-noise limit.}
\end{center}
\end{figure}
\end{Suppfig}
%%------------------

\subsection{Hardware noise}
Noise from hardware used for the noise model.
\begin{table}[h]
\begin{ruledtabular}
\begin{tabular}{c c c c c c c c}
Qubit & \shortstack{T1 \\ (us)} & \shortstack{T2 \\ (us)}  & 
\shortstack{Readout \\ assignment error}&
\shortstack{Prob meas0 \\ prep1} &
\shortstack{Prob meas1 \\ prep0} &
\shortstack{ID/RZ/ \\ X error} &
\shortstack{CX error \\ (pair: error)} \\
\hline
\centering 1 & 0.2363e-03 & 0.3716e-03 & 0.0106 & 0.0120 & 0.0092 & 0.4087e-03 & 1-2: 8.4263e-03 \\
\hline
\centering 2 & 0.3084e-03 & 0.3122e-03 & 0.0282 & 0.0304 & 0.0260  & 0.1757e-03 & 2-3: 8.9020e-03 \\
\hline
\centering 3 & 0.3654e-03 & 0.2516e-03 & 0.0387 & 0.0280 & 0.0494  & 0.2400e-03 & 3-4: 4.9061e-02 \\
\hline
\centering 4 & 0.2834e-03 & 0.1213e-03 & 0.1763 & 0.1734 & 0.1792  & 0.3426e-03 & 4-3: 4.9061e-02 \\
\end{tabular}
\end{ruledtabular}
\caption{\label{tab:hardware_parameters}
Noise model from the hardware used for the noiseless simulator for computing the ground state with zero-noise extrapolation.
}
\end{table}

\end{document}